# Quest for an efficient mathematical and computational method to explore optimal extreme weather modification

Yohei Sawada[1]

[1] Department of Civil Engineering, Graduate School of Engineering, the University of Tokyo, Tokyo, Japan

Corresponding author: Y. Sawada, Department of Civil Engineering, the University of Tokyo, Tokyo, Japan, 5-3-1, Hongo, Bunkyo-ku, Tokyo, Japan, yoheisawada@g.ecc.u-tokyo.ac.jp

**Abstract**

It is a grand challenge to find a feasible weather modification method to mitigate the impact of extreme weather events such as tropical cyclones. Previous works have proposed potentially effective actuators and assessed their capabilities to achieve weather modification objectives through numerical simulations. However, few studies have explored efficient mathematical and computational methods to inversely determine optimal actuators from specific modification goals. Here I demonstrate the utility of the ensemble Kalman filter (EnKF)-based control method, referred to as ensemble Kalman control (EnKC). The series of numerical experiments with the Lorenz 96 model indicates that EnKC efficiently identifies local, small, and intermittent control perturbations that can mitigate extreme events. The existing techniques of EnKF, such as background error covariance localization and observation error covariance inflation, can improve the sparsity and efficiency of the control. This work paves the way toward the real-world applications of EnKC to explore the controllability of extreme atmospheric events.

# 1. Introduction

Artificial modification of extreme weather conditions, such as Tropical Cyclones (TCs), is a grand challenge in meteorology. For instance, Project STORMFURY attempted to mitigate the maximum wind speeds of TCs by artificially stimulating convection outside the eyewall through cloud seeding, although the effectiveness of this approach has come into question recently (Willoughby et al. 1985). Cotton et al. (2007) proposed the injection of dust into TCs inspired by numerical simulation performed by Zhang et al. (2007), which showed that Saharan dust substantially affects TC development. Latham et al. (2012) explored the potential of cooling sea surface temperature to weaken TCs. Jacobson et al. (2014) investigated the ability of large offshore wind turbines to reduce peak near-surface TC wind speeds. A comprehensive review of interventions in severe weather is available in Miller et al. (2023).

In these previous works, various actuators (e.g., aerosols, dust, and turbines) were first identified, and their potential to achieve weather modification goals (e.g., significant reduction of maximum wind speed) was assessed by numerical simulations. It is also strongly necessary to develop mathematical and computational methods to inversely identify optimal actuators from the modification objectives. Considering that humanity can apply extremely local, small, and intermittent forces to Earth systems and the evaluation of weather modification is carried out by computationally expensive Earth system models, it is crucial to develop computationally efficient methods to find local and small perturbations that can significantly alter the future evolution of weather conditions. Henderson et al. (2005) tackled this challenge for the first time. They fully leveraged the four-dimensional variational method (4DVAR), a data assimilation method of numerical weather prediction, to identify the optimal perturbation to mitigate the damage caused by strong winds in TCs.

While Henderson et al. (2005) demonstrated the effectiveness of their mathematical method assuming that the intervention could be performed only once, Miyoshi and Sun (2022) proposed the frequent addition of small perturbations and the adaptive change in directions of the perturbations. Their proposed framework of idealized experiments, known as the Control Simulation Experiment (CSE), has recently been adopted in other works (Sun et al. 2023; Ouyang et al. 2023; Kawasaki and Kotsuki 2024). However, previous works in CSE, as well as the work by Henderson et al. (2005), have not fully explored how to adaptively obtain local and small perturbations in a computationally efficient manner which can be applied to computationally expensive atmospheric models.

Recently, Sawada (2024) proposed the ensemble Kalman filter (EnKF)-based control method to efficiently identify optimal perturbations to control spatio-temporal chaos. In the present paper, the proposed EnKF-based method is called ensemble Kalman control (EnKC). EnKC fully uses the existing EnKF architecture which has been successfully used for state estimation of Earth systems. EnKC is an iteration-free and derivative-free method. EnKC inherently minimizes the same cost function as model predictive control, allowing it to efficiently explore minimum perturbations necessary to achieve the desired future state using model predictions. Although Sawada (2024) demonstrated that EnKC could effectively control the Lorenz 63 system (Lorenz 1963), it has not yet been shown how EnKC can enforce perturbation added to the system remain local and small. Here I demonstrate that existing techniques of EnKF to improve state estimation and prediction, specifically background error covariance localization and observation error covariance inflation, can be used to identify local and small perturbations. Toward real-world applications, this paper discusses how to use EnKC to explore local and small perturbations to mitigate extreme events in spatio-temporal chaos.

## 2. Method
### 2.1. Ensemble Kalman Filter (EnKF)
A discrete state-space dynamic system is defined as:

$$\boldsymbol{x}_t = M(\boldsymbol{x}_{t-1}) + \boldsymbol{q}_{t-1} \tag{1}$$

$$\boldsymbol{y}_t^o = H(\boldsymbol{x}_t) + \boldsymbol{r}_t \tag{2}$$

where $\boldsymbol{x}_t$ is the state variables at time $t$, $M$ demotes the dynamic model, $\boldsymbol{q}_t$ is the noise process representing model error at time $t$, $\boldsymbol{y}_t^o$ is the observation at time $t$, $H$ is the observation operator, and $\boldsymbol{r}_t$ is the noise process representing observation error at time $t$. EnKF adjusts the model-estimated state variables by minimizing the following cost function:

$$J(x_t) = \frac{1}{2}\left(\boldsymbol{x}_t - \overline{\boldsymbol{x}_t^b}\right)^T \boldsymbol{P}^{b^{-1}}\left(\boldsymbol{x}_t - \overline{\boldsymbol{x}_t^b}\right) + \frac{1}{2}\left(\boldsymbol{y}_t^o - H(\boldsymbol{x}_t)\right)^T \boldsymbol{R}^{-1}\left(\boldsymbol{y}_t^o - H(\boldsymbol{x}_t)\right) \tag{3}$$

where $\overline{\boldsymbol{x}_t^b}$ is the background ensemble mean of state estimates from the ensemble, $\boldsymbol{P}^b$ is the background error covariance matrix estimated from ensemble simulation, and $\boldsymbol{R}$ is the observation error covariance matrix.

In a pure Monte Carlo implementation, the solution to this minimization is transporting the $i$th member of an $N$-member ensemble by the following equations:

$$x_t^{a(i)} = x_t^{f(i)} + K\left(y_t^o - H\left(x_t^{f(i)}\right)\right) \tag{4}$$

$$K = P^b H^T (H P^b H^T + R)^{-1} \tag{5}$$

where $x_t^{a(i)}$ and $x_t^{f(i)}$ is the state variables of the *i*th member of analysis and forecast ensemble, respectively, $K$ is the Kalman gain, $H$ is the linearized observation operator. In this paper, the stochastic filter (Burgers et al. 1998; Houtekamer and Mitchell 1998) was used, so that a randomly perturbed vector of $y_t^o$ was used to update ensemble members. In the implementation of EnKF, the linearized observation operator is unnecessary (although this paper uses the linear observation operator) since the following ensemble-based approximations of $P^b H$ and $H P^b H^T$ are used:

$$P^b H^T = \frac{1}{N-1} \sum_{i=1}^{N} (x_t^{f(i)} - \overline{x_t^b})\left(H\left(x_t^{f(i)}\right) - \overline{H\left(x_t^{f(i)}\right)}\right)^T \tag{6}$$

$$H P^b H^T = \frac{1}{N-1} \sum_{i=1}^{N} (H\left(x_t^{f(i)}\right) - \overline{H\left(x_t^{f(i)}\right)})\left(H\left(x_t^{f(i)}\right) - \overline{H\left(x_t^{f(i)}\right)}\right)^T \tag{7}$$

where $\overline{x_t^b} = \frac{1}{N}\sum_{i=1}^{N} x_t^{f(i)}$, $\overline{H\left(x_t^{f(i)}\right)} = \frac{1}{N}\sum_{i=1}^{N} H\left(x_t^{f(i)}\right)$. This approximation also ensures that a full covariance matrix of $P^b$ is unnecessary to compute EnKF updates.

Given the limited ensemble sizes, covariance localization is necessary to mitigate the impact of sampling error on the quality of EnKF updates. A smooth correlation function $\rho$ has been applied to the equation of Kalman gain:

$$K = \rho \circ P^b H^T (\rho \circ H P^b H^T + R)^{-1} \tag{8}$$

where the symbol $\circ$ denotes the Schur product. In this paper, $\rho$ is defined as:

$$\rho = \exp\left(-\frac{d(i,j)}{L}\right) \tag{9}$$

where $d(i,j)$ shows the distance between model grid point $i$ and observation $j$, and $L$ is the scale parameter. See Houtekamer and Zhang (2016) for the comprehensive review of EnKF.

## 2.2. Ensemble Kalman Control (EnKC)

After obtaining the analysis ensemble, $x_t^{a(i)}$, an extended ensemble forecast is performed from the analysis ensemble members over the control horizon, $T_c$. Then, the following minimization problem is solved to identify the optimal perturbation that moves the system state to the desired position:

$$J^c(x_t) = \frac{1}{2}(x_t - \overline{x_t^a})^T P^{a-1}(x_t - \overline{x_t^a}) + \frac{1}{2}(r_{t+T_c} - H^c(x_{t+T_c}))^T R_c^{-1}(r_{t+T_c} - H^c(x_{t+T_c})) \quad (10)$$

$$s.t. x_{t+1} = M(x_t)$$

where $\overline{x_t^a}$ is the analysis ensemble mean, $P^a$ is the analysis error covariance matrix, $r_t$ is the reference vector indicating the desired state at time $t$, $H^c$ is the operator to map state variables onto the reference vector to evaluate how the estimated future state meets control criteria, and $R_c$ is the user-defined weights. Note that minimization of equation (10) seeks the smallest perturbation, $x_t - \overline{x_t^a}$, to be added to the initial state to effectively minimize the difference between the future state and the control objective, based on model predictions. This concept is fully analogous to model predictive control (see Schwenzer et al. 2021 for a comprehensive review). Discussion of the similarities between existing data assimilation algorithms in geoscience and model predictive control can be found in Sawada (2024).

Assuming that the dynamics in the control horizon is linear, minimizing (10) can be approximately done by the ensemble Kalman smoother (EnKS):

$$x_t^c = \overline{x_t^a} + K\left(r_{t+T_c} - \overline{H^c\left(x_{t+T_c}^{a(i)}\right)}\right) \quad (11)$$

$$K = P^{a,c} \mathbf{H}^{cT} \left(\mathbf{H}^c P^c \mathbf{H}^{cT} + R_c\right)^{-1} \quad (12)$$

where $x_t^c - \overline{x_t^a}$ is the estimated perturbation to be added to the system at time $t$. $x_{t+T_c}^{a(i)}$ is the model state variables at time $t + T_c$ predicted from the analysis state variables as initial conditions. $\mathbf{H}^c$ is the linearized operator of $H^c$. $P^c$ is the error covariance estimated by ensemble-estimated state variables at time $t + T_c$. $P^{a,c}$ is the cross-covariance between analysis ensemble at time t and predicted ensemble at time $t + T_c$. As demonstrated in equations (6)-(7), the Kalman gain can be approximated as:

$$P^{a,c} \mathbf{H}^{cT} = \frac{1}{N-1} \sum_{i=1}^{N} (x_t^{a(i)} - \overline{x_t^a}) \left(H\left(x_{t+T_c}^{a(i)}\right) - \overline{H\left(x_{t+T_c}^{a(i)}\right)}\right)^T \quad (13)$$

$$\mathbf{H}^c P^b \mathbf{H}^{cT} = \frac{1}{N-1} \sum_{i=1}^{N} (H\left(x_{t+T_c}^{a(i)}\right) - \overline{H\left(x_{t+T_c}^{a(i)}\right)}) \left(H\left(x_{t+T_c}^{a(i)}\right) - \overline{H\left(x_{t+T_c}^{a(i)}\right)}\right)^T \quad (14)$$

where $\overline{x_t^a} = \frac{1}{N}\sum_{i=1}^{N} x_t^{a(i)}$, $\overline{H(x_t^{a(i)})} = \frac{1}{N}\sum_{i=1}^{N} H(x_t^{a(i)})$. As Sawada (2024) found, the equations (11)-(14) indicate that the minimization of Equation (10) can be achieved by projecting the model-based ensemble prediction onto the control criteria and assimilating $r_{t+T_c}$ as a "pseudo-observation" vector with a "pseudo observation error covariance" $R_c$, into the analysis state variables using EnKS. Note that only $x_t^c - \overline{x_t^a}$ is necessary to control the system, so that it is unnecessary to update each ensemble member in the step of EnKC.

The algorithm of EnKC can be outlined as follows (see also Sawada (2024)):
Step 1: Perform EnKF using the initial guess (i.e. background ensemble) and real observations to estimate the state variables at time $t$.
Step 2: Using the analysis ensemble obtained from Step 1, perform ensemble forecasting up to time $t + T_c$. Project the predicted state variables at $t + T_c$ onto control criteria using the operator $H^c$.
Step 3: Perform EnKS, using the analysis ensemble and reference vector, $r_{t+T_c}$ as pseudo-observations. By assimilating $r_{t+T_c}$, obtain the appropriate perturbation to be added to a system, $x_t^c - \overline{x_t^a}$ as a solution of ETKS.
Step 4: Add the perturbation $x_t^c - \overline{x_t^a}$ obtained in Step 3 to the real nature. The same perturbation is also added to all analysis ensemble members to accurately estimate the modified nature.
Step 5: With the updated analysis ensemble, perform ensemble prediction up to time $t + T$ (note that $T$ is the duration of the data assimilation window). This prediction serves as the new initial guess for EnKF. Then, return to Step 1.

**2.3. Techniques to make control perturbation local and small**
The control perturbation estimated by EnKC described in Section 2.2, $x_t^c - \overline{x_t^a}$, has the same dimension as the state variables of the system. The vanilla EnKC explained in Section 2.2 requires perturbing all state variables in every control step, which is apparently unrealistic in the context of weather modification. Therefore, it is necessary to enforce sparsity on $x_t^c - \overline{x_t^a}$. To enforce sparsity on control perturbations, the cost function should be modified as follows:

$$J^c(x_t) = \frac{1}{2}(x_t - \overline{x_t^a})^T P^{a-1}(x_t - \overline{x_t^a}) + \frac{1}{2}(r_{t+T_c} - H^c(x_{t+T_c}))^T R_c^{-1}(r_{t+T_c} - H^c(x_{t+T_c})) + \lambda|x_t - \overline{x_t^a}|_{l_0} \quad (15)$$

where $|.|_{l_0}$ denotes the number of non-zero entries of the vector and $\lambda$ is a parameter. Although Schneider et al. (2022) proposed the algorithm to minimize (15) using ensemble Kalman inversion, in which model parameters are estimated by iteratively applying EnKS updates, it is difficult to directly apply their method to our problem setting due to its high computational cost. Instead of directly minimizing (15), this paper proposes two empirical methods to enforce sparsity on control perturbations.

First, covariance localization is again used for EnKC. The Kalman gain in EnKC (Equation 12) was modified as:

$$\mathbf{K} = \rho_c \circ \mathbf{P}^{a,c} \mathbf{H}^{cT} \left( \rho_c \circ \mathbf{H}^c \mathbf{P}^c \mathbf{H}^{cT} + \mathbf{R}_c \right)^{-1} \tag{16}$$

which is fully analogous to Equation (8) in EnKF. The localization function in EnKC, $\rho_c$, is defined as:

$$\rho_c = \begin{cases} 1 & if\ d(i,j) < L_c \\ 0 & otherwise \end{cases} \tag{17}$$

where $d(i,j)$ is the distance between model grid point $i$ and location of the reference vector (i.e., $r_{t+T_c})j$, and $L_c$ is a parameter. It is assumed that the "location" of the control objective can be defined in this method. In extreme weather modification, the control objective is preventing extreme state variables in a specific area, so that the control objective can reasonably be assigned to some model grids in most cases. Equations (16) and (17) empirically assumed that the control perturbations should be added near the location where an extreme event occurs, and computed perturbations far from the event can be neglected.

The second method to enforce sparsity on control perturbations is inspired by Schneider et al. (2022). After the vanilla EnKC described in Section 2.2, the following function is applied to all entries of $x_t^c - \overline{x_t^a}$:

$$\mathrm{T}(\theta) = \begin{cases} 0 & if\ |\theta| < \sqrt{2\lambda} \\ \theta & otherwise \end{cases} \tag{18}$$

which simply assumes that small entries of control perturbation vectors are noise and/or unimportant. While Schneider et al. (2022) recognized $\sqrt{2\lambda}$ as a fixed parameter in their $\lambda$-thresholding algorithm, this study adaptively changes the threshold according to the maximum entries of $x_t^c - \overline{x_t^a}$:

$$\sqrt{2\lambda} = \Lambda * \max(|x_t^c - \overline{x_t^a}|) \tag{19}$$

where $\Lambda$ is a parameter to control the enforced sparsity. In Equation (19), max(.) picks up the largest absolute value of all entries of the vectors. A larger $\Lambda$ forces more entries of control perturbation vectors to zero. When $\Lambda = 1$, a single grid which has the largest entry of the EnKC-estimated control perturbation is perturbed.

These proposed methods can reduce the number of non-zero entries in $x_t^c - \overline{x_t^a}$, which makes the estimated control perturbations more realistic. There are two other benefits of these methods in enforcing sparsity on $x_t^c - \overline{x_t^a}$. First, as discussed in Section 2.1, the negative impact of sampling error in the covariance matrix on the estimation of control perturbations can be mitigated. Second, these methods can eliminate extremely small perturbations which realistic actuators cannot implement considering their signal-to-noise ratio. Even if flapping of butterfly wings could influence cyclones, it is beyond human capability to control them.

In addition to removing small and noisy patterns of control perturbations, it is absolutely necessary to prevent the generation of excessively large perturbations since realistic actuators cannot introduce such large perturbations into the atmosphere. EnKC minimizes Equation (10) and intrinsically has a mechanism to find the small perturbations under the error covariances of $P^a$ and $R_c$. If $R_c$ is set to relatively larger than $P^a$, the obtained perturbations become smaller, which is fully analogous to EnKF updates. However, large $R_c$ places less weight on meeting the control criteria, making it difficult for the system's state to stay in the desired conditions. This is also fully analogous to EnKF updates for state estimation, where a large $R_c$ reduces the analysis increments and places more weight on initial guess. In this paper, I analyze the tradeoff between control performance and the magnitude of perturbations by changing $R_c$.

Even with large $R_c$, a large perturbation is generated when the difference between the estimated and targeted control reference, $r_{t+T_c} - \overline{H^c\left(x_{t+T_c}^{a(i)}\right)}$, is large. A similar problem exists in the ordinary EnKF for state estimation. When the innovation, $y_t^o - H\left(x_t^{f(i)}\right)$, is large, unrealistically large analysis increments are generated, which can drive state variables far from the system's attractor and degrade the prediction skill. To address this issue, Minamide and Zhang (2017) proposed Adaptive Observation Error Inflation (AOEI). In AOEI, the observation error covariance is adaptively determined by the following equation:

$$\sigma_o^2 = max\left\{\sigma_{ot}^2, \left(y_t^o - \overline{H\left(x_t^{f(i)}\right)}\right)^2 - \sigma^2_{H\left(x_t^{f(i)}\right)}\right\} \qquad (20)$$

where $\sigma_o^2$ is the observation error variance, $\sigma_{ot}^2$ is the flow-independent observation error variance, $\sigma^2_{H\left(x_t^{f(i)}\right)}$ is the ensemble-estimated variance of an observable variable.

See Minamide and Zhang (2017) for the complete description of AOEI and its effectiveness. Note that observation error correlations were neglected in this study as many EnKF works did, so that this paper only focuses on the observation error variance. I used AOEI to prevent from adding extremely large perturbations:

$$\sigma_{oc}^2 = max\left\{\sigma_{oct}^2, \left(r_{t+T_c} - \overline{H^c\left(x_{t+T_c}^{a(i)}\right)}\right)^2 - \sigma^2_{H^c\left(x_{t+T_c}^{a(i)}\right)}\right\} \quad (21)$$

where $\sigma_{oc}^2$ is the diagonal part of $R_c$, and $\sigma_{oct}^2$ is the flow-independent part of $\sigma_{oc}^2$. When the ensemble mean of the extended forecast from the analysis ensemble is far from the control objective, $R_c$ is inflated to avoid applying excessively large control forces.

### 3. Experiment design

Inspired by Sun et al. (2023), I conducted a CSE aimed at mitigating extreme values of the Lorenz 96 system (Lorenz 1995). The Lorenz 96 system is described by:

$$\frac{dX_k}{dt} = (X_{k+1} - X_{k-2})X_{k-1} - X_k + F \quad (22)$$

for $k \in \{1,2,\ldots\ldots,K\}$ and $K = 40$ in this study. $F$ was set to 8.0. The system has periodic boundary conditions, so that $X_0 = X_k$, $X_{-1} = X_{k-1}$, and $X_{k+1} = X_1$. Equation (22) was solved by the 4th order Runge-Kutta method with the timestep of 0.05.

It is assumed that the system can be observed in grid points with even grid numbers (20 state variables can be observed). The observation frequency was set to 0.05. Observations were generated by adding the Gaussian white noise, whose mean and variance were 0 and 1.0, respectively, to the nature run. This observation error was assumed to be known, and EnKF was performed every 0.05 timestep with the localization parameter $L = 2.0$ (see equation (9)).

The objective of the control is to mitigate extremely large positive values in the Lorenz 96 system. An extended forecast was performed after EnKF. The control horizon, $T_c$, was set to 0.2. I performed EnKC only when the ensemble mean of the extended forecast at $t + T_c$ indicated $X_k > 12$ at more than one grid point. It is commonly accepted that 0.2 unit of time in the Lorenz 96 model corresponds to one day in reality, and $X_k = 12$ corresponds to an approximately 99.9 percentile extreme event. It can be reasonably assumed that society has a consensus to perform weather modification, which potentially has a negative impact on environments, when the majority of forecast ensemble members indicates extreme events in one day. Since the goal is to prevent the state variables from further increasing when $X_k > 12$, it is logical to set the control objective, $r_{t+T_c}$ to 12. In EnKC, the pseudo-observation of $r_{t+T_c} = 12$ was assimilated into the system using EnKS. This pseudo-observation is assigned to the grid point where $X_k > 12$.

I tested the EnKF and EnKC methods described in Section 2 with ensemble sizes of 40, 20, and 10. Initially, I did not use AOEI and set the fixed standard deviation of the diagonal part of $R_c$ to 0.0001, 0.001, 0.01, 0.1, and 1.0 (note that the off-diagonal part of $R_c$ was 0). When the covariance localization was used to enforce sparsity on control perturbations (equation (16)-(17)), a parameter in equation (17), $L_c$, was set to 1, 2, 5, and 10. When the ignorance of small non-zero entries of control perturbation was used to enforce sparsity on control perturbation (equation (18)-(19)), a parameter in equation (19), $\Lambda$, was set to 0.25, 0.5, 0.75, and 1.0. To evaluate the effectiveness of these approaches to enforce sparsity on control perturbations, I randomly chose $n_L$ entries of control perturbation vectors and made the other $40 - n_L$ entries zero in every EnKC step as an additional experiment. The number of the randomly selected entries, $n_L$, was set to 1, 3, 9, and 19. I also performed the EnKC experiments in which I did not enforce sparsity on control perturbation (i.e. modifying state variables at all grid points). I repeated the EnKC experiments with the same parameter set applying AOEI. In the case of the AOEI experiments, the fixed standard deviation of the diagonal part of $R_c$ described above was used as the minimal observation error standard deviation ($\sigma_{oct}$ in equation (21)).

I performed the numerical simulations with 160600 timesteps. In the first 14600 timesteps, EnKC was not performed as a spin-up. In this spin-up period, only EnKF is performed to synchronize the estimated state variables with the uncontrolled nature run. Then, both EnKF and EnKC were performed in the remaining 146000 timesteps. All results in this 146000 timesteps were used to evaluate the performances.

## 4. Results

Figure 1 demonstrates the impact of the control method on extreme events in the Lorenz 96 system. Here, the control weight ($R_c$) was set to 0.1, and EnKC ignored the small entries of control perturbations with $\Lambda = 0.5$. The ensemble size was set to 40. Figure 1b shows a significant reduction of the magnitude of 99.999 percentile extreme state variables, indicating that EnKC successfully mitigates the extreme events. Figure 1a illustrates that interventions from EnKC do not alter the climatology of the Lorenz 96 system, which is strongly crucial for real-world applications. This modification is achieved by relatively local, small, and intermittent external perturbations (Figure 2).

The difference of the 99.999 percentile state values between uncontrolled nature and controlled nature is defined as a performance indicator of the intervention. Figures 3a and 3d show that larger control weights (smaller $\boldsymbol{R}_c$) and larger localization scales (larger $L_c$ or smaller $\Lambda$) yield better performance, revealing a clear trade-off between the impact of the interventions and the magnitude of the interventions. Since EnKC with $\Lambda$ of 1.0, 0.75, 0.5, and 0.25 modifies the state variables at 1, 2, 4, and 10 grids on average, respectively, Figures 3a and 3d indicate that local perturbations imposed at fewer than 3 grid points can significantly mitigate the 99.999 percentile severe events. This performance found in the experiments with a 40-ensemble ensemble size substantially degrades with smaller ensemble sizes (Figures 3b, 3c, 3e, and 3f). There are two reasons for this degraded performance with smaller ensemble sizes. First, due to the sampling error of $\boldsymbol{P}^b$, state estimation by EnKF suffers from the small ensemble size, which degrades the accuracy of the extended forecast in the control horizon. Second, even if the state estimation were accurate, the small ensemble size leads to erroneous estimation of $\boldsymbol{P}^c$ and $\boldsymbol{P}^{a,c}$, degrading the accuracy of the control perturbation. Figures 3g-3l show that AOEI reduces the overall performance, which is reasonable since AOEI decreases the magnitude of control perturbations as discussed later. A distinctive characteristic of AOEI is that it can reduce the dependence of the control performances on ensemble sizes. When the ensemble size is small, the ensemble forecast tends to be over-confident (i.e. ensemble spreads are too small), leading to the biased estimation of control perturbations. In this case, AOEI strongly inflates $\boldsymbol{R}_c$ (see equation (21)) and mitigates the negative effect of sampling errors on the estimation of control perturbations. The performances shown above could not be obtained when randomly choosing the controlled points (Figures 3m-o), so that the proposed methods to enforce sparsity on control perturbations can effectively make the control perturbation local without substantially degrading the control performance.

In addition to the locality of the control perturbations, their magnitude is evaluated. Figure 4 shows the averaged maximum entry of $\left|\boldsymbol{x}_t^c - \overline{\boldsymbol{x}_t^a}\right|$ during the experiment period. Since the maximum perturbation that can be added at one model grid point is restricted by the specifications of actuators, it is important to verify whether this maximum entry is reasonably small. The maximum perturbation substantially depends on the control weight, $R_c$. If ones need to make the maximum perturbation smaller, larger $R_c$ should be chosen. Compared with the covariance localization, the method of ignoring of small perturbations to enforce sparsity of control perturbations results in larger maximum perturbations, as this method always pick up the maximum entry of the original $|x_t^c - \overline{x_t^a}|$. AOEI substantially reduces the maximum control perturbations. In AOEI, when the innovation, $r_{t+T_c} - \overline{H^c\left(x_{t+T_c}^{a(i)}\right)}$, becomes larger, $R_c$ increases applying smaller control weights. This mechanism contributes to setting smaller upper bounds on the magnitude of control perturbations. Considering the relatively small differences in performance between EnKC with and without AOEI especially when the localization scale is small, AOEI effectively reduces the required maximum magnitude of control perturbations. It should be noted that smaller ensemble sizes lead to larger maximum perturbations, reducing the efficiency of the control.

Figure 5 shows the L2-norm of the control perturbation vectors, which may be correlated with the total energy required to add the control perturbations. It is intuitively apparent that the total energy increases with larger localization scales and greater control weights (smaller $R_c$). AOEI reduces the total energy required from EnKC. Figure 6 also highlights the potential benefits of AOEI. While EnKC without AOEI has many outliers and large interquartile ranges of the estimated total energy, EnKC with AOEI consistently generates small perturbations. This characteristic of AOEI is potentially useful for designing reasonable actuators, although the overall reduction of extreme state variables is smaller as shown in Figure 3. Note that EnKC leverages the non-linear growth of the added perturbations although it assumes linear system's dynamics. The larger reduction in extreme states than the magnitude of added perturbations is achieved. For instance, when control weights are set to minimal ($R_c = 1.0$), the mean L2-norms of the control perturbation vectors estimated by EnKC with AOEI, which neglects smaller perturbations, are less than 0.16. In these cases, the achieved reduction of the 99.999$^{\text{th}}$ percentile extreme state variables ranges from approximately 0.2 to 0.6, depending on the different localization scales. It should also be noted that the magnitude of the control perturbation increases with smaller ensemble sizes. Since the performance of the control method is reduced by smaller ensemble sizes, as shown in Figure 3, the efficiency of the control method substantially decreases when insufficient ensemble sizes are used.

Figures 5 and 7 show that the magnitude of the control perturbations is inversely correlated with the frequency of adding perturbations. When the added perturbation is small, it is sometimes necessary to perform multiple interventions. The frequency of interventions increases with smaller ensemble sizes. This occurs because the degradation of the skill of the extended forecast within the control horizon leads to perturbations added at the inappropriate times.

## 5. Discussion and Conclusions

In this paper, I proposed a mathematical and computational method to identify optimal actuators of weather modification based on prescribed objectives. In the context of weather modification, the proposed method should efficiently find local, small, and intermittent control perturbations. The proposed EnKC (Sawada 2024) can be the useful iterative-free and derivative-free method to identify flow-dependent optimal perturbations. I found that the existing techniques of EnKF, such as background covariance localization and observation covariance inflation, can contribute to improving the sparsity and efficiency of the control. The proof-of-concept numerical experiment with the Lorenz 96 model shows promise toward the real-world applications of EnKC for exploring the controllability of extreme atmospheric events.

Although mathematical and computational methods to identify optimal actuators for weather modification have been investigated, the proposed EnKC method and the findings of this paper significantly differ from earlier works. Henderson et al. (2005) used the widely-used data assimilation architecture, 4DVAR, to find the optimal perturbation to mitigate tropical cyclones. The advantages of the EnKC method over Henderson et al. (2005) are: (1) the ability to adaptively estimate control perturbations using flow-dependent background covariance information, (2) the avoidance of iterative calculation and the tangent-linear model, and (3) the demonstration that existing background covariance localization and observation covariance inflation methods of EnKF can effectively be used to find local and small perturbations. Miyoshi and Sun (2022) and Sun et al. (2023) realized to adaptively estimate control perturbations. However, their ensemble-based control method, which uses the vector of the difference between two best and worst ensemble members based on a control criterion to generate perturbations, lacks mechanisms to objectively determine the magnitude of perturbations. I propose a flexible and objective method to adaptively determine the magnitude and direction of perturbations based on the existing theory of EnKF. As Kawasaki and Kotsuki (2024) demonstrated, the direct application of model predictive control can provide accurate control perturbations under complex control objectives and constraints. However, model predictive control in control engineering is computationally expensive and challenging to apply to Earth system scientific problems in which numerical models are complex and computationally expensive. One can fully rely on the existing architecture of computationally efficient EnKFs to implement control simulation experiments by EnKC, making the proposed approach readily applicable to real-world weather modification problems. The major limitation of EnKC compared to previous works is that EnKC assumed that the dynamics within the control horizon are linear, so that it is necessary to frequently perform EnKF and EnKC to avoid the nonlinear error growth. Recently, extremely frequent satellite and radar observations are available and have been used for EnKF (e.g., Sawada et al. 2019; Honda et al. 2018; Miyoshi et al. 2016). EnKC should leverage such frequent updates to explore appropriate control perturbations in relatively short control horizons.

Future works should focus on applying EnKC to atmospheric models to modify severe weather events such as tropical cyclones and sudden heavy rainfall. It remains a grand challenge to confirm that severe weather can be significantly modified by perturbations which are local, small, and intermittent enough to be implemented by mankind. As demonstrated in this paper, this challenge will be addressed by investigating the trade-off between the performance of the control methods and the cost of the control perturbations quantified by locality, magnitude, and frequency based on the results of EnKC applications to atmospheric simulations.

**Acknowledgements**
This work was supported by the JST Moonshot R&D program (Grant JMPJMS2281).

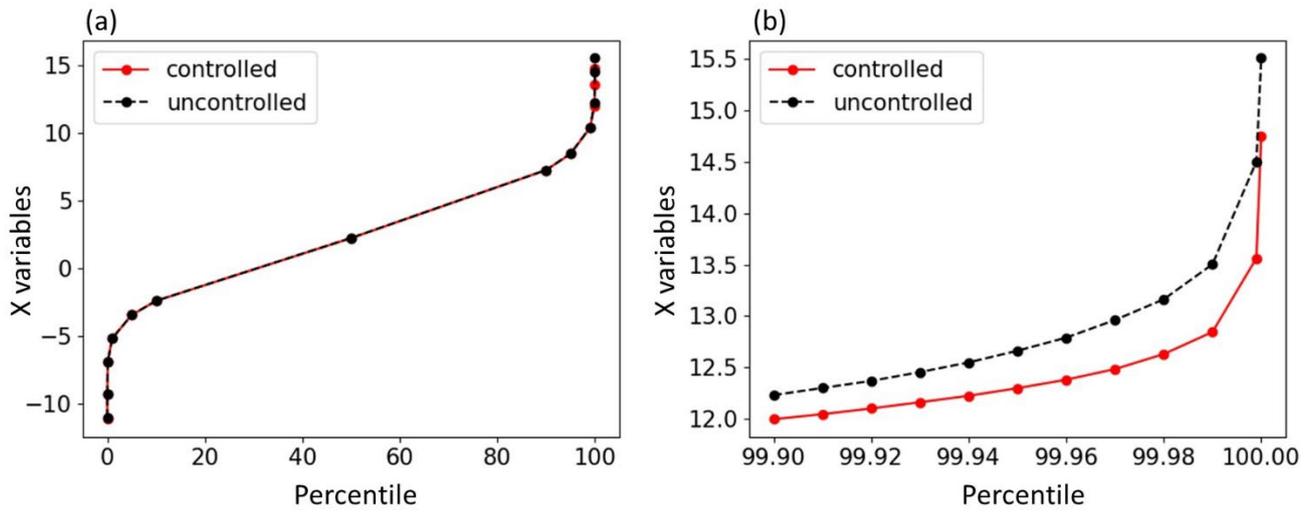

**Figure 1.** The distribution of state variables of the Lorenz 96 system in the control simulation experiments. Red and black dashed lines are controlled and uncontrolled nature runs, respectively. While (a) shows the whole percentile ranges from 0 to 100, (b) shows the zoom-in of the range from 99.9 percentile to the maximum. In this experiment, the ensemble size is set to 40, the diagonal entry of $\boldsymbol{R_c}$ is set to 0.1, and $\Lambda = 0.5$ (see also the manuscript).

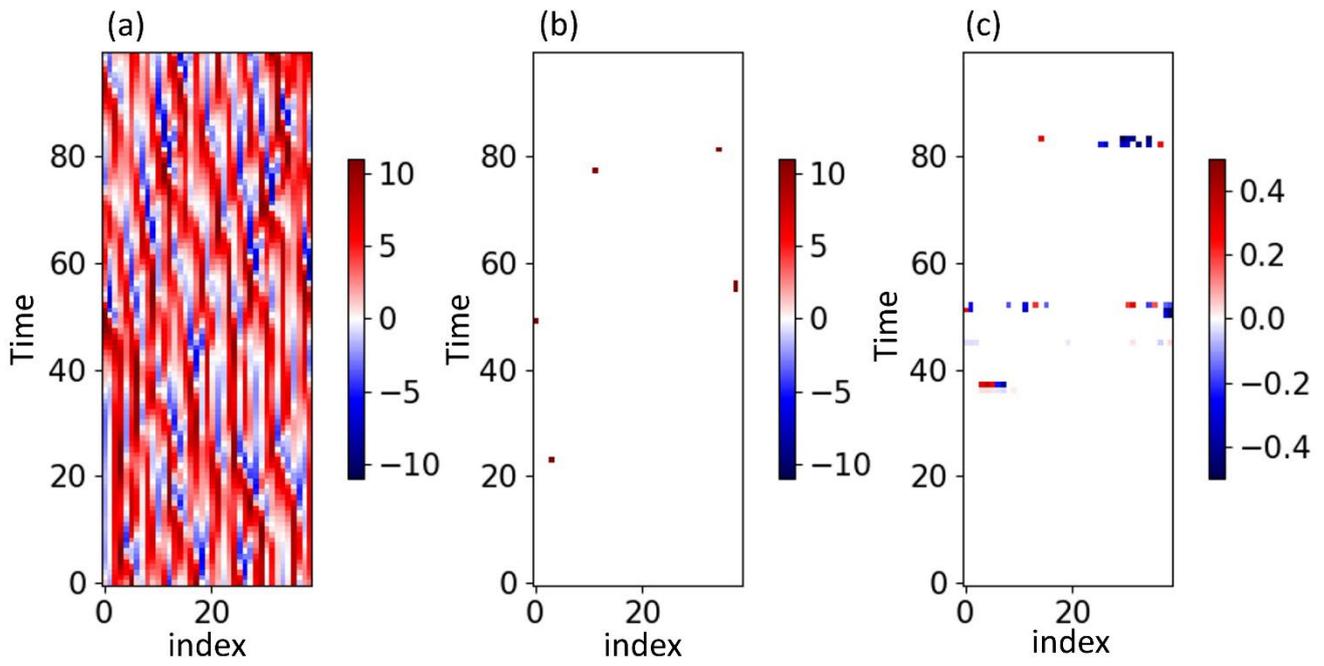

**Figure 2.** (a) Hovmöller diagram of $X_k$ in the Lorenz 96 system. (b) same as (a), but only for $X_k > 12$. (c) EnKC-estimated control perturbations. In this experiment, the ensemble size is set to 40, the diagonal entry of $\mathbf{R_c}$ is set to 0.1, and $\Lambda = 0.5$ (see also the manuscript).

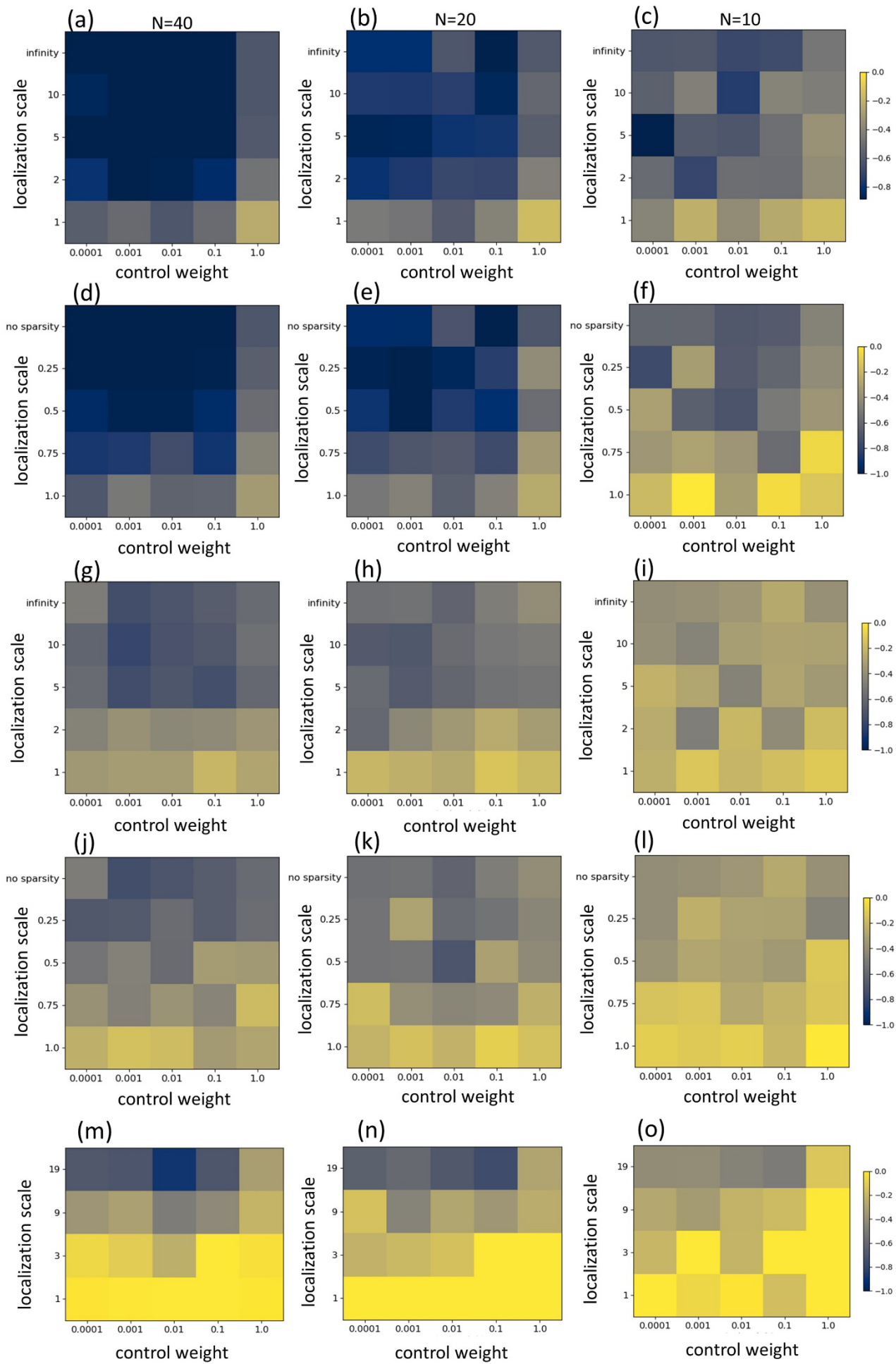

**Figure 3.** The difference of 99.999 percentile state values between uncontrolled nature and controlled nature. (a-c) the EnKC with covariance localization without AOEI. (d-f) the EnKC with the ignorance of small entries of control purturbations without AOEI. (g-i) the EnKC with covariance localization with AOEI. (j-l) The EnKC with the ignorance of small entries of control purturbations with AOEI. (m-o) the EnKC with the random selection of perturbed entries without AOEI. The ensemble size is set to 40 (a,d,g,j,m), 20 (b,e,h,k,n), and 10 (c,f,i,l,o). Horizontal axis and vertical axis of each panel show control weight ($R_c$ or $\sigma_{oct}$) and localization scale ($L_c$, $\Lambda$, or $n_L$), respectively.

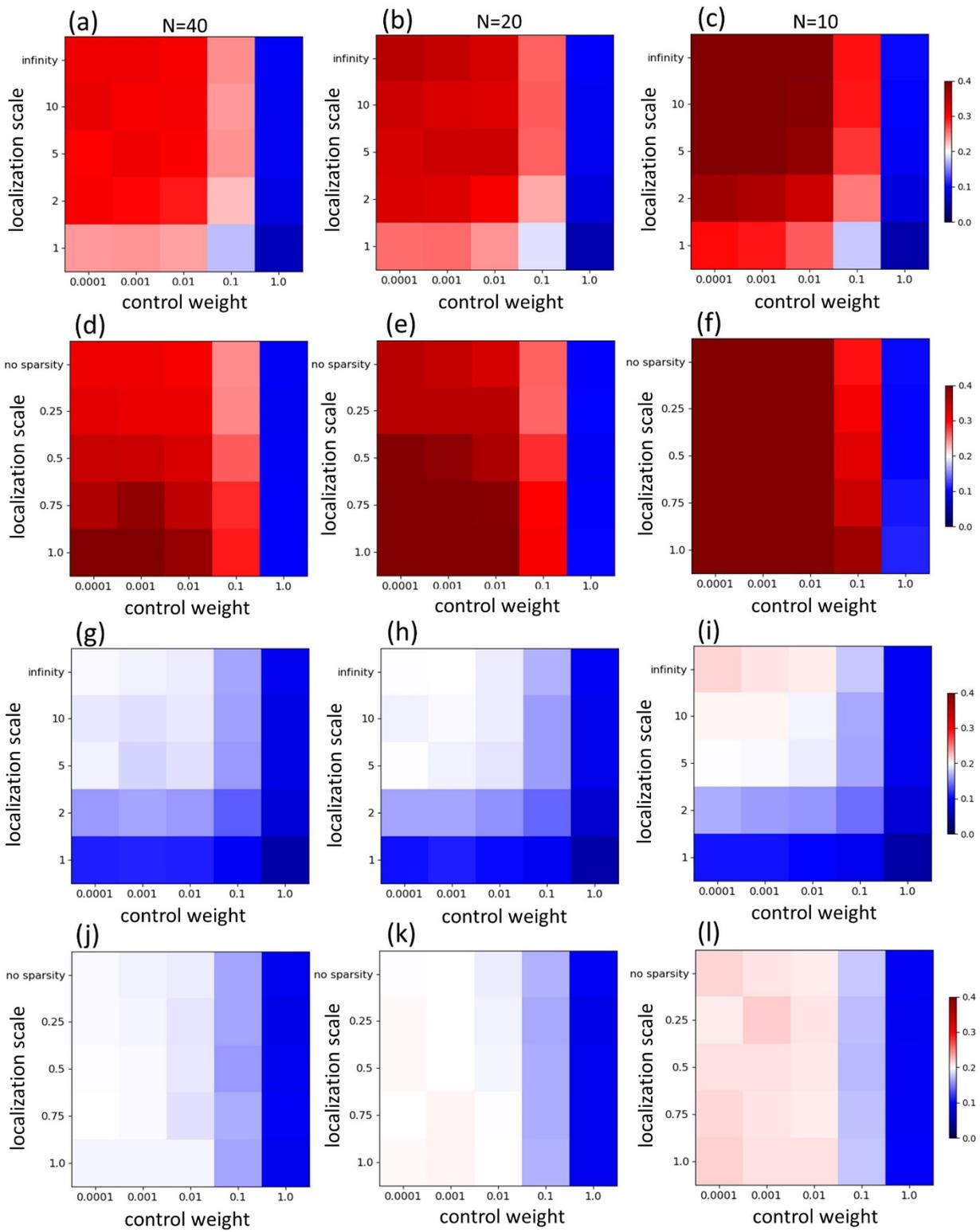

**Figure 4.** Same as Figures 3a-3l but for the averaged maximum entries of control perturbation vectors.

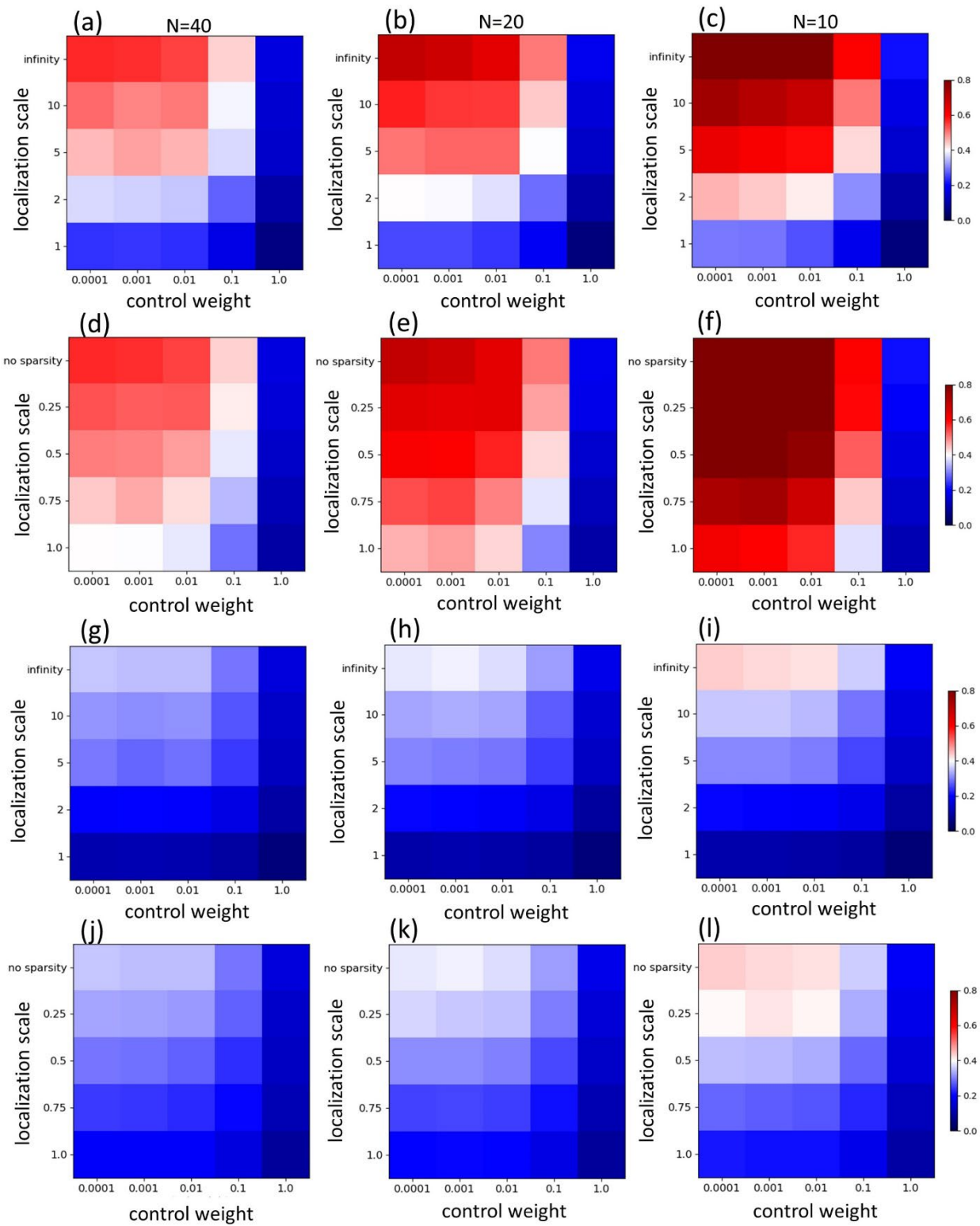

**Figure 5.** Same as Figures 3a-3l but for the averaged L2-norm of control perturbation vectors.

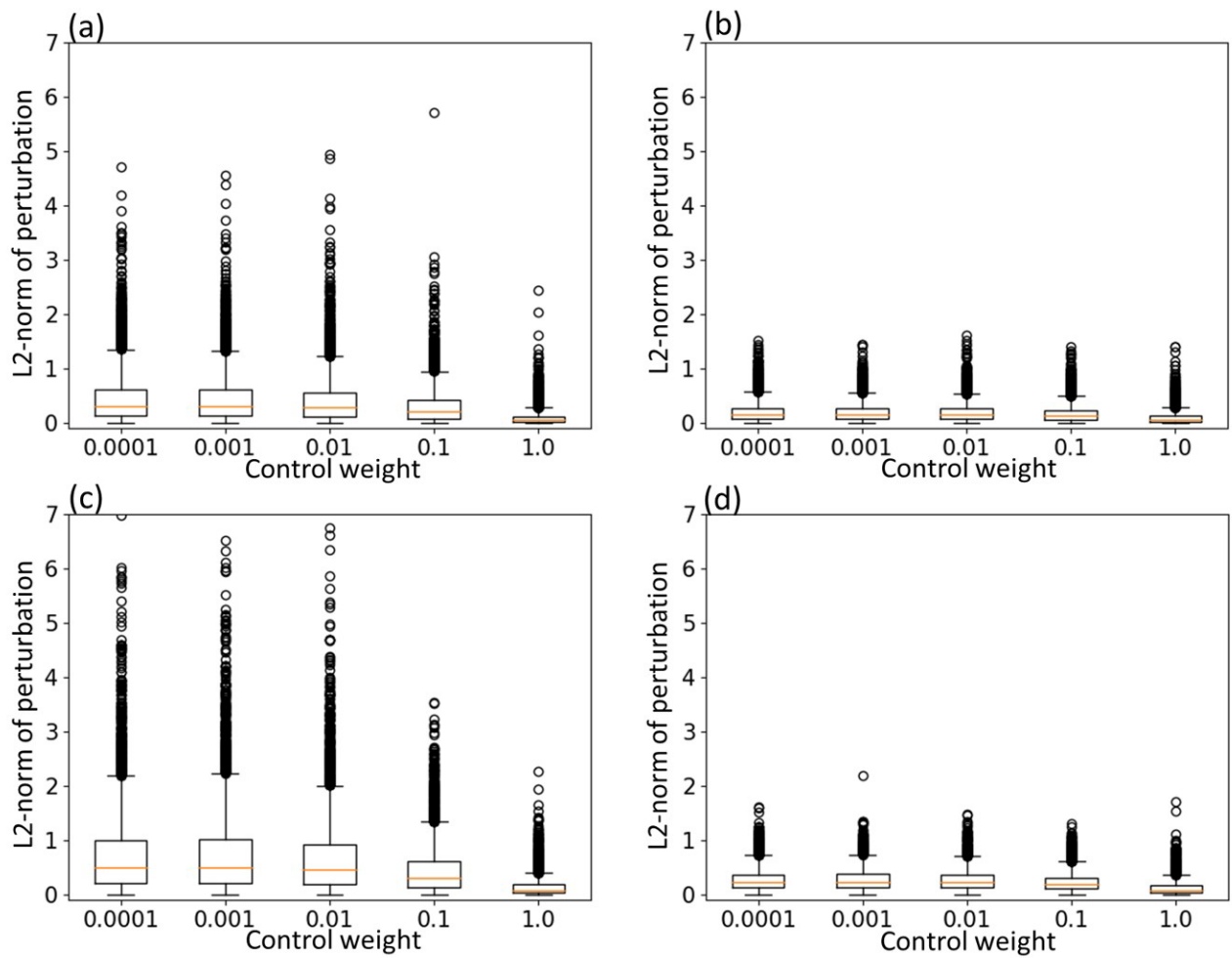

**Figure 6.** Boxplots of the L2-norms of control perturbation vectors estimated by EnKC which has (a) covariance localization and no AOEI, (b) covariance localization and AOEI, (c) ignorance of small entries of control perturbations and no AOEI, and (d) ignorance of small entries of control perturbations and AOEI. In (a)-(b), the localization scale, $L_c$, is 3. In (c)-(d), $\Lambda = 0.75$.

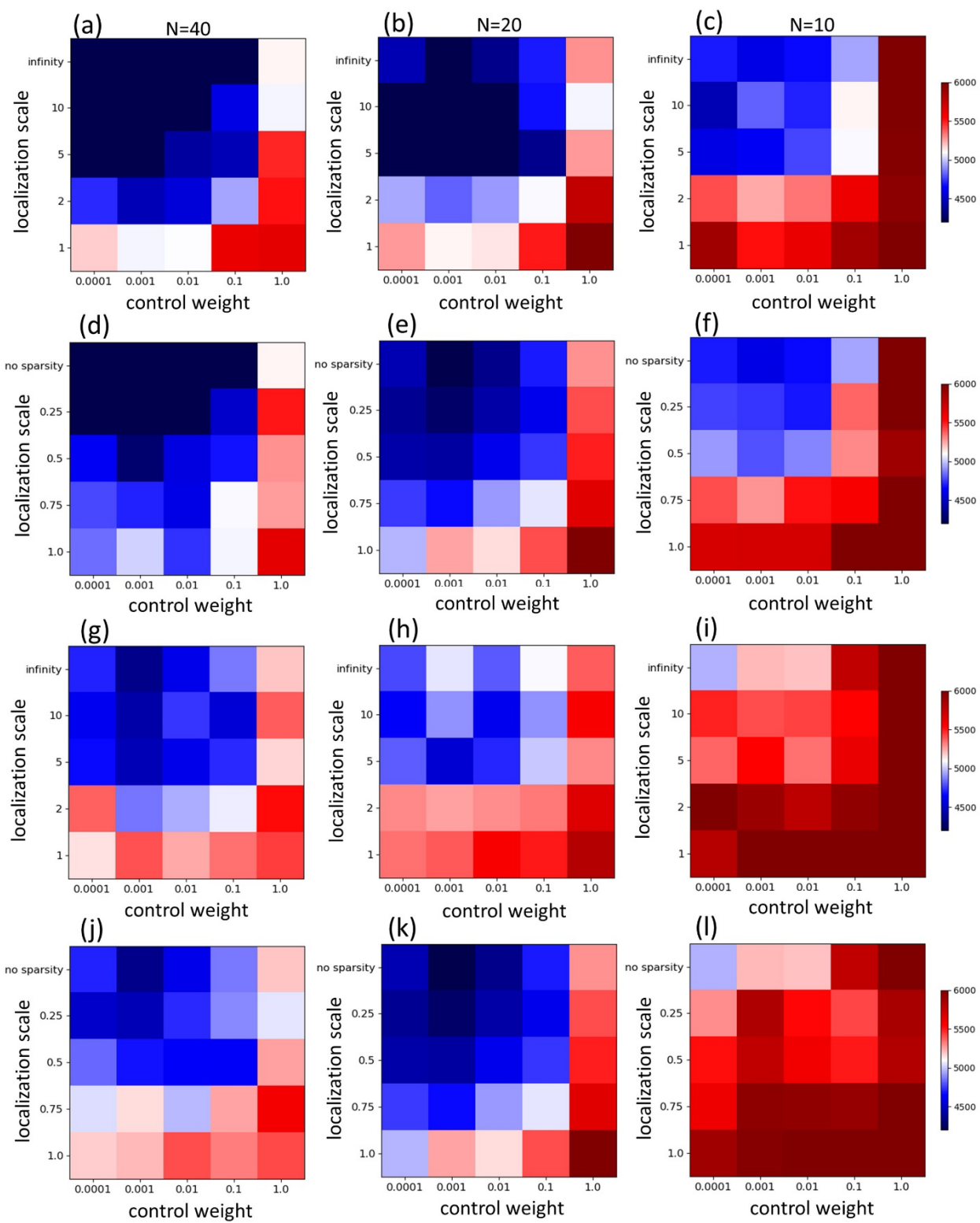

**Figure 7.** Same as Figures 3a-3l but for the frequencies of interventions.